\documentstyle[12pt,epsfig]{article}
\newcommand{\be}[1]{\begin{equation} \label{(#1)}}
\newcommand{\ee}{\end{equation}}
\newcommand{\ba}[1]{\begin{eqnarray} \label{(#1)}}
\newcommand{\ea}{\end{eqnarray}}
\newcommand{\nn}{\nonumber}
\newcommand{\rf}[1]{(\ref{(#1)})}
\def \lsim {\mbox{${}^< \hspace*{-7pt} _\sim$}}
\def \gsim {\mbox{${}^> \hspace*{-7pt} _\sim$}}

\def\t{{\tau}^-\rightarrow l^{\pm} \pi^{\mp}\pi^-}
\def\tlnv{{\tau}^-\rightarrow l^{+} \pi^{-}\pi^-}
\def\tlfv{{\tau}^-\rightarrow l^{-} \pi^{+}\pi^-}
\def \lg  {\langle}
\def \rg  {\rangle}
\begin{document}
\begin{flushright}               
\hfill{ USM-TH-104}               
\end{flushright}                 
\vskip1cm                        
\begin{center}
   {\Large\bf Sterile neutrinos in tau lepton decays}\\[3mm]
Vladimir Gribanov$\ ^1$, Sergey Kovalenko
\footnote{On leave from
the Joint Institute for Nuclear Research, Dubna, Russia}
and  Ivan Schmidt\\[1mm]
{\it Departamento de F\'\i sica, Universidad
T\'ecnica Federico Santa Mar\'\i a,}\\
{\it Casilla 110-V, Valpara\'\i so, Chile}
\end{center}
\bigskip

\begin{abstract}
We study possible contributions of heavy sterile neutrinos $\nu_h$ 
to the decays $\tau^-\rightarrow e^{\pm}(\mu^{\pm})\pi^{\mp}\pi^-$. From the experimental 
upper bounds on their rates we derive new constraints on the $\nu_h-\nu_{\tau}$ mixing 
in the mass region $140.5\mbox{ MeV}\leq m_{\nu_h}\leq 1637$ MeV. We discuss 
cosmological and astrophysical status of $\nu_h$ in this mass region and compare our
constraints with those recently derived by the NOMAD collaboration.
\end{abstract}
\vskip 0.5cm

\bigskip
\bigskip


\bigskip
\bigskip


\newpage

\section{Introduction}

Hypothetical sterile neutrinos $\nu_s$, blind to the electroweak interactions, 
were invoked into particle physics in various contexts as  possible means to 
resolve observed anomalies. The most prominent case deals with neutrino 
anomalies: solar neutrino deficit, atmospheric neutrino anomaly and 
results of the LSND neutrino experiment. Simultaneous explanation of all these three 
anomalies in terms of neutrino oscillations requires at least one sterile neutrino $\nu_s$
\cite{nu-rev}. 
In this case $\nu_s$ together with $\nu_{e,\mu,\tau}$ allows one to introduce three 
independent mass square differences associated with these three anomalies. 
Although the recent Super Kamiokande global analysis \cite{SK-global} disfavours 
active-sterile neutrino oscillations as a dominant channel this hypothesis 
is not ruled out.  
Along this line one may reasonably admit the existence of more than one sterile neutrino,
say, one per generation, as it is suggested by various extensions of the SM.
In this situation among the neutrino mass eigenstates, which are superpositions of 
the active $\nu_{e,\mu,\tau}$ and sterile $\nu_{s(i)}$ weak eigenstates,  one can 
encounter not only light neutrinos, 
but also heavy states $\nu_h$. The question of viability of 
this scenario is the subject of experimental searches as well as cosmological 
and astrophysical constrains.

Recently some phenomenological, cosmological and astrophysical issues of 
the intermediate mass neutrinos $\nu_h$ in the MeV mass region have been addressed in 
the literature \cite{Dolgov}. This was stimulated by the attempts of explanation of 
the KARMEN anomaly \cite{KARMEN} in terms of a neutrino state with a mass of 33.9 MeV 
mixed with $\nu_{\tau}$  \cite{BPS:95}. Although recent data of this collaboration 
\cite{KARMEN2} have not confirmed this anomaly, the question of existence of heavy sterile 
neutrinos $\nu_h$ remains open.

These sterile neutrinos can be searched for  
as peaks in differential rates of various processes and by direct production of $\nu_h$ 
followed by their decays in the detector 
(for summary see Ref. \cite{PDG} p. 361.).
The $\nu_h$ can also give rise to significant enhancement of the total rate of certain 
processes if their masses happen to be located in an appropriate region \cite{DGKS:2000}. 
This effect would be especially pronounced in reactions that are forbidden in the SM. 
The lepton number/flavor violating (LNV/LFV) processes belong to this category. Many of them are 
stringently restricted by experiment and allow one to derive stringent limits on 
the $\nu_h$ contribution.

In the present paper we study the $\nu_h$ contribution to the LFV and LNV $\tau$-decays:  
$\tau^-\rightarrow e^{-}(\mu^-) \pi^+ \pi^-$ and  
$\tau^-\rightarrow e^{+}(\mu^+) \pi^- \pi^-$. The first (LFV) process can receive contribution 
both from Dirac and Majorana neutrinos while to the second (LNV) process only Majorana 
neutrinos can contribute.
These processes are capable to provide us with unique information on the $\nu_h-\nu_{\tau}$ 
mixing matrix element $U_{\tau h}$. 
In the mass region $140.5\mbox{GeV}\leq m_{h} \leq 1637\mbox{MeV}$, which we are going to study, 
the $\nu_h$ contribution to the considered $\tau$-decays gains resonant enhancement
\cite{DGKS:2000}.
This effect makes discussed $\tau$-decays very sensitive to presence of $\nu_h$ with the masses
$m_h$ in the resonant region.
Under certain assumptions we extract from the experimental data \cite{CLEO} on these $\tau$-decays 
new constraints on the $U_{\tau h}$ matrix element in the resonant 
region for both Majorana and Dirac heavy sterile neutrinos 
$\nu_h$. Up to recently there were no credible laboratory limits on this quantity in the MeV 
mass region \cite{PDG}. Only recent data of NOMAD collaboration \cite{NOMAD} allowed 
establishing constraints on $U_{\tau h}$ in the MeV mass region.  
In that part of the resonant region, which overlaps with the region probed by NOMAD, 
our constraints are more stringent by around two orders of magnitude.

\section{Neutrino mass matrix and interactions}

Consider an extension of the SM with the three left-handed weak
doublet neutrinos $\nu'_{Li} = (\nu'_{Le},\nu'_{L\mu},\nu'_{L\tau})$
and $n$ species of the SM singlet right-handed neutrinos
$\nu'_{Ri}=(\nu'_{R1},...\nu'_{Rn})$.
The general mass term for this set of fields can be written as
\ba{mass-term}
\nn
&-& \frac{1}{2} \overline{\nu^{\prime}} {\cal M}^{(\nu)} \nu^{\prime c} +
\mbox{H.c.} =
- \frac{1}{2}
(\bar\nu'_{_L},  \overline{\nu_{_R}^{\prime c}})
\left(\begin{array}{cc}
{\cal M}_L & {\cal M}_D \\
{\cal M}^T_D  & {\cal M}_R \end{array}\right)
\left(\begin{array}{c}
\nu_{_L}^{\prime c} \\
\nu'_{_R}\end{array}\right) + \mbox{H.c.} =\\
&-&\frac{1}{2} \sum_{i=1}^{3+n} m_{\nu i} \overline{\nu^c}_{i}\nu_i
+ \mbox{H.c.}
\ea
Here ${\cal M}_L, {\cal M}_R$ are $3\times 3$ and $n\times n$ symmetric
Majorana mass matrices, ${M}_D$ is $3\times n$ Dirac type matrix.
Rotating the neutrino mass matrix by the unitary transformation to the diagonal form 
\ba{rotation}
U^T {\cal M}^{(\nu)}U = Diag\{m_{\nu i}\}
\ea
we end up with $n+3$ Majorana neutrinos
$\nu_i =  U^*_{ki} \nu'_{k}$ with the masses $m_{\nu i}$.
In special cases there may
appear among them pairs with masses degenerate in
absolute values. Each of these pairs can be collected into a Dirac neutrino
field. This situation corresponds to conservation of certain lepton numbers
assigned to these Dirac fields.

The considered generic model must contain at least three observable light neutrinos
while the other states may be of arbitrary mass.  In particular, they may include
hundred MeV neutrinos $\nu_h$, which we are going to consider in the next section. 
Presence or absence of these neutrino states is a question for experimental searches.

In this scenario neutrino mass eigenstates have in general non-diagonal LFV couplings 
to the Z-boson
\ba{Z-width}
Z^{\mu} \sum_{\alpha = e,\mu,\tau}
\overline{\nu'_{\alpha}} \gamma_{\mu} P_L \nu'_{\alpha} =
Z^{\mu} \sum_{\alpha = e,\mu,\tau}\sum_{i,j=1}^{n+3}
U_{\alpha i}U^*_{\alpha j}
\overline{\nu_{j}} \gamma_{\mu} P_L \nu_{i} \equiv
\sum_{i,j=1}^{n+3} {\eta}_{ij}
\overline{\nu_{j}} \gamma_{\mu} P_L \nu_{i},
\ea
where the last two expressions are written in the mass eigenstate basis. For the case
of only three massive neutrinos one has  ${\cal P}_{mn}=\delta_{mn}$
as a consequence of unitarity of $U_{\alpha n}$. In general
${\cal P}_{mn}$ is not a diagonal matrix and flavor changing neutral currents 
in the neutrino sector become possible at tree level.

As is known the invisible Z-boson width $\Gamma_{inv}^Z$ is an efficient counter of the 
number of neutrinos. In the introduced scenario, despite possible presence of an 
arbitrary number of light neutrinos, the predicted value of the 
$\Gamma_{inv}^Z$ is always consistent with its experimental value.
Indeed, if all the neutrinos are light with masses $m_{\nu i}<<M_Z$ their contribution 
to the invisible Z-boson width can be written as
\ba{inv-width} 
\Gamma_{inv}^Z = \sum_{i,j=1}^{n+3} \left |{\eta}_{ij}\right |^2 
\Gamma_{\nu}^{SM} = \Gamma_{\nu}^{SM} \sum_{\alpha,\beta = e,\mu,\tau} 
\delta_{\alpha\beta}\delta_{\alpha\beta} = 3\Gamma_{\nu}^{SM}. 
\ea 
With the SM prediction for the partial Z-decay width to a pair of light neutrinos
$\Gamma_{\nu}^{SM} = 167.24 \pm 0.08$ MeV this formula always reproduces the experimental 
value of $\Gamma_{inv}^Z = 498.8\pm 1.5$ MeV. The chain of equalities in Eq. \rf{inv-width} 
follows again from  the unitarity of $U_{\alpha n}$. Thus, independently of 
the number of light neutrinos with masses $m_{\nu}<< M_Z$ the factor 3 in 
the last step counts the number of weak doublet neutrinos. 
This conclusion is changed in the presence of heavy neutrinos $N$ with 
masses $M_N > M_Z/2$ which do not contribute to $\Gamma_{inv}$. In this 
case the unitarity condition is no longer valid and the factor 3 is 
changed to a smaller value. 

Having these arguments in mind we introduce in the next section 
neutrino states $\nu_h$ with masses in the hundred MeV region.
These states can be composed of sterile and active neutrino flavors
as described in the present section.

\section{$\tau^-\rightarrow e^{\pm}(\mu^{\pm})\pi^{\mp}\pi^-$ decay rates}
Neutrinos contribute to the LNV $\tau$-decay $\tlnv$ 
according to the lowest order diagrams shown in Fig.1.
The $\nu$-contribution to the LFV $\tau$-decay $\tlfv$ is determined by
the tree-level diagram similar to the diagram in Fig. 1(a). The loop-diagram 
analogous to the diagram in Fig. 1(b) is absent in the latter case.
\begin{figure}[h!]
\mbox{ \epsfxsize=16 cm\epsffile{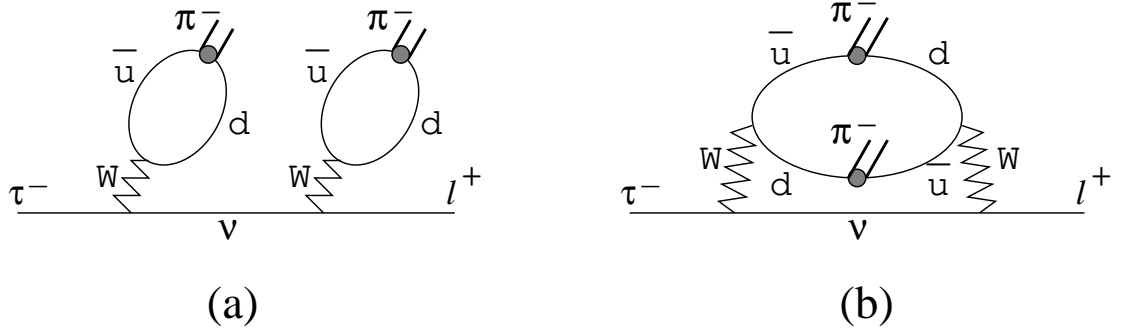}}
\vspace{-12cm}
\caption{The lowest order diagrams contributing to 
$\tau^-\rightarrow l^{+} \pi^{+} \pi^-$ decay.
}
\vskip 1cm
\end{figure}
The diagram  in Fig. 1(b)
requires knowledge of the $\pi$-meson wave function at long distances. As yet it is 
poorly known and would introduce into the calculations uncontrollable uncertainties. 
In the present paper we concentrate on the resonant mass domain determined in \rf{domain}, 
where the diagram in Fig. 1(a) absolutely dominates over that in Fig. 1(b). 
For this reason we 
neglect this diagram latter on. 
The hadronic structure parameters necessary for 
calculation of the tree-level contribution Fig. 1(a) involves only one parameter of 
hadronic structure, the pion decay constant $f_{\pi}$, accurately known from
experiment. For the $\t$ decay
with $l = e, \mu$ we derive the following decay rate formula
\ba{rate-lnv}
 && \Gamma(\tlnv)=
 c
\int\limits_{s_{l}^-}^{s^+} d s \left | \sum_{k}  \frac{U_{l
k}U_{\tau k} m_{\nu k}}{s - m_{\nu k}^2}\right |^2 G^{l
}(\frac{s}{m_{\tau}^2}) +
  \nn \\&&2
\frac{c}{m_{\tau}^{2}}{\rm Re}[\int\limits_{s_l^-}^{s^+} d s
\left(\sum_k \frac{U_{l k}U_{\tau k} m_{\nu k}}{s - m_{\nu k}^2}\right)
\int\limits_{v_{l}^-}^{v_{l}^+} d v \left(\sum_n\frac{U_{l n}U_{\tau n}
m_{\nu k}}{v - m_{\nu n}^2}\right)^* H^{l}(\frac{s}{m_{\tau}^2},\frac{v}{m_{\tau}^2})],\\
\label{rate-lfv}
 && \Gamma(\tlfv)=
 c  \int\limits_{s_{l}^-}^{s^+} d s  \left | \sum_{k} \frac{U_{lk}^*U_{\tau k}}
 {s - m_{\nu k}^2}\right |^2 s\  G^{l}(\frac{s}{m_{\tau}^2}).
 \ea
The unitary mixing matrix $U_{ij}$ relates $\nu'_i = U_{ij}\nu_h$
weak $\nu'$ and mass $\nu$ neutrino eigenstates. The numerical
constant is $c = (G_F^4/32) (\pi)^{-3}
f^4_{\pi}   m^5_{\tau} |V_{ud}|^4$, where  $f_{_\pi} = 93$MeV 
and $m_\tau = 1777$MeV is the $\tau$-lepton mass.
We introduced the functions
\ba{G-funct}\nn
&&G^{l}(z) = \phi^{l}(z) z^{-2}\times 
  [(z-x_l^2)^2 -x_{\pi}^2(z+x_{l}^2)]
 [(z-1)^2 -x_{\pi}^2(z+1)], \\ 
&&H^{l}(z_1,z_2)= (1+x_l^2-z_1-z_2)(x_l^2 - z_1z_2) 
+ \\ \nn
&& + x_\pi^2 \left[(1+x_l^2)(z_1+z_2-x_\pi^2)-
 2 (z_1z_2 + x_l^2)\right],
 \ea
where $x_{i} = m_i/m_{\tau}$, $\lambda(x,y,z) =
x^2+y^2+z^2-2xy -2yz- 2xz$, $\phi^{l}(z) = 
\lambda^{1/2}(1,z,x_{\pi}^2)\lambda^{1/2}(z,x_{\pi}^2,x_l^2)$. 
The integration limits in Eqs. \rf{rate-lnv}, (\ref{rate-lfv}) are
 \ba{lim-int}
&&s_l^- = m_{\tau}^2(x_{\pi} + x_{l})^2, \ \ \ \ s^+ =
m_{\tau}^2(1 - x_{\pi})^2,\\ \nn &&v_{l}^{\pm} =\frac
{m_{\tau}^2}{2y}
\left[(x_{\pi}^2-x_{l}^2)(x_{\pi}^2-1)+y(1+x_{l}^2+2x_{\pi}^2)-
y^2\pm \phi^{l}(y) \right] \ea with $y=s/m_{\tau}^2$.

Assuming that there exist only light neutrinos 
with masses $m_{\nu i}<< m_{\tau}$ we can rewrite Eqs. \rf{rate-lnv}, (\ref{rate-lfv})
approximately in the form 
\ba{small-mass}
\Gamma(\tlnv)=  c\ \frac{|\lg m_{\nu}\rg_{l \tau}|^2}{m_{\tau}^2} {\cal A}_{+}^{(l)},\ \ \ 
\Gamma(\tlfv)=  c\  \frac{|\lg m^2_{\nu}\rg_{l \tau}|^2}{m_{\tau}^4} {\cal A}_{-}^{(l)},
\ea
where 
\ba{coeff}
\nn
{\cal A}_{+}^{(l)} &=&
 m^2_{\tau}\int\limits_{s_{l}^-}^{s^+} \frac{d s}{s^2}\ G^{l}(\frac{s}{m_{\tau}^2}) +
2 \int\limits_{s_l^-}^{s^+} \frac{d s}{s}
\int\limits_{v_{l}^-}^{v_{l}^+} \frac{d v}{v}\ H^{l}(\frac{s}{m_{\tau}^2},\frac{v}{m_{\tau}^2}),\\
{\cal A}_{-}^{(l)} &=&
   m^4_{\tau} \int\limits_{s_{l}^-}^{s^+} \frac{d s}{s^3}\  G^{l}(\frac{s}{m_{\tau}^2})
\ea
and 
\ba{eff-m}
\lg m_{\nu}\rg_{l \tau} =  \sum_{k} U_{l k} U_{\tau k}\ m_{\nu k},\ \ \
\lg m^2_{\nu}\rg_{l \tau} =  \sum_{k} U^*_{l k} U_{\tau k}\ m^2_{\nu k}.
\ea
From atmospheric and solar neutrino oscillation data, combined with the tritium 
beta decay endpoint, one can derive upper bounds on masses of all the three 
neutrinos \cite{barg1} $m_{e,\mu,\tau}\leq 3$ eV. Then we have conservatively \cite{DGKS:NANPino}   
\ba{estim}
|\lg m_{\nu}\rg_{\tau l}| < 9\ \mbox{eV},\ \ \
|\lg m^2_{\nu}\rg_{\tau l}| < 27\ \mbox{eV}^2. 
\ea
With these upper bounds we obtain from Eqs. \rf{small-mass} a rough estimate 
\ba{rough}
&&R_{e^+(\mu^+)}\equiv \frac{\Gamma(\tau^-\rightarrow e^+(\mu^+)\pi^{-}\pi^-)}
{\Gamma(\tau^-\rightarrow All)}\  \lsim\  3.9(2.6)\times 10^{-31},\\
&&R_{e^-(\mu^-)}\equiv 
\frac{\Gamma(\tau^-\rightarrow e^-(\mu^-)\pi^{+}\pi^-)}
{\Gamma(\tau^-\rightarrow All)}\  \lsim\  3.0(1.2)\times 10^{-47}.
\ea
This is to be compared with the present experimental bounds on these branching ratios \cite{PDG}
\ba{tdec-lim}
  {\cal R}_{e^+(\mu^+)}^{exp}\leq 1.9(3.4)\times 10^{-6}, \ \ \
 {\cal R}_{e^-(\mu^-)}^{exp}\leq 2.2(8.2)\times 10^{-6}, \ \ \
\ea
The comparison clearly shows that the light neutrino contributions 
to the processes $\tau^-\rightarrow l^{\pm}\pi^{\mp}\pi^-$
are far from being detected in the near future. On the other hand
experimental observation of $\t$ decays at larger rates would
indicate some new physics beyond the SM, or the presence of an extra neutrino 
state $\nu_h$ with the mass in the hundred MeV domain.

Assume that there exist neutrinos $\nu_h$ with masses $m_h$ 
in the interval
\ba{domain}
 \sqrt{s_e^-} \approx 140.5 \mbox{MeV} \leq m_{h} \leq
\sqrt{s^+} \approx 1637 \mbox{MeV}. 
\ea
These neutrinos would give resonant contributions to the processes
$\t$ since the first term in Eqs. \rf{rate-lnv} and the expression in 
Eq. (\ref{rate-lfv}) have non-integrable singularities at $s=m_h^2$. 
Apparently, in the resonant region \rf{domain} one has to take into account 
the finite width $\Gamma_{\nu h}$ of the neutrino $\nu_h$. This 
is accomplished  by the substitution 
$m_{h}\rightarrow m_{h} - (i/2)\Gamma_{\nu h}$ in Eqs. \rf{rate-lnv}, (\ref{rate-lfv}).
As we will show in the next section the total decay width of the neutrino is small 
$\Gamma_{\nu h}<< m_{\nu h}$. Therefore in the resonant domain \rf{domain} the neutrino 
propagator in Eqs. \rf{rate-lnv}, (\ref{rate-lfv}) 
has a very sharp maximum at $s=m_{h}^2$. The second
term in Eq. \rf{rate-lnv}, being finite in the limit $\Gamma_{\nu h} =0$, can be
neglected in this case. This allows us  to  rewrite Eqs. \rf{rate-lnv}, (\ref{rate-lfv}) 
in the resonant mass region with good precision in a simple form  
\ba{estim3}
 \Gamma^{res}(\tau^-\rightarrow l^{\pm}\pi^{\mp}\pi^-)& \approx& c
\pi G^{l}(z_0)\frac{m_h |U_{\tau h}|^2 |U_{l h}|^2} { \Gamma_{\nu h}},
\ea
 with $z_0 = (m_h/m_\tau)^2$.

\section{Heavy sterile neutrino decays}
Heavy sterile neutrinos $\nu_h$ can decay in both charged (CC) 
and neutral (NC) current channels 
$\nu_h \rightarrow l_1 l_2 \nu_2; l M(q_1\bar{q_2})$ and  
$\nu_h \rightarrow \nu_i l \bar l; \nu_i M(q\bar{q})$, where $l=e,\mu$ and $M$ are mesons
specified below. These decays are induced by the Lagrangian terms
\ba{CC-NC}
{\cal L} = \frac{g_2}{\sqrt{2}}\  U_{l i}\ \bar l \gamma^{\mu} P_L \nu_i\  W^-_{\mu}
+  \frac{g_2}{2 \cos\theta_W}\  U_{\alpha j} U_{\alpha i}^*\  \bar\nu_i \gamma^{\mu} P_L \nu_j\  
Z_{\mu},
\ea
where $l = e,\mu,\tau$ and $U_{\alpha\beta}$ is the neutrino mixing matrix 
defined in Eq. \rf{rotation}. Despite the fact that the NC term is of second 
order in the mixing matrix the NC decay channels of the heavy sterile neutrinos $\nu_h$, 
as will be seen, are as important as the CC one.

Calculating the total width $\Gamma_{\nu h}$ we divide the interval 
\rf{domain} into two parts 
\ba{parts}
\sqrt{s_e^-} \approx 140.5 \mbox{MeV} \leq &m_{h}& \leq  m_{\eta'}=\ \ \ 958 \mbox{MeV}
\ \ \ \   \mbox{domain (I)}, \\ \nn
m_{\eta'}=958 \mbox{MeV} < &m_h& \leq\sqrt{s^+} \approx 1637 \mbox{MeV}\ \ \ \ \mbox{domain (II)},
\ea
where $m_{\eta'} = 958$ MeV is the mass of the isoscalar pseudo-scalar meson $\eta'(958)$.

In the domain (I) we obtain $\Gamma_{\nu h}$ directly, 
by calculating all the $\nu_h$ decay channels to the leptons $l$
and mesons $M$ one by one, using the following meson matrix elements
\ba{hadron}
&& \langle\pi^-(p)|\bar d \gamma_\mu \gamma_5 u|0\rangle =i \sqrt{2} f_\pi
 p_\mu, \ \ \ 
 \langle K^-(p)|\bar s \gamma_\mu \gamma_5 u|0\rangle=-i \sqrt{2} f_K
 p_\mu, \\ \nn
&& \langle\rho^-(p)|\bar d \gamma_\mu \gamma_5 u|0\rangle = \sqrt{2} m_\rho
 f_\rho  p_\mu \epsilon_\mu^*,\ \ \ 
 \langle K^{*-}(p)|\bar s \gamma_\mu \gamma_5 u|0\rangle = m_{K^*}
 f_{K^*}  p_\mu \epsilon_\mu^*,\\ \nn
&&  \langle\pi^0(p)|\bar u \gamma_\mu \gamma_5 u|0\rangle = 
- \langle\pi^0(p)|\bar d \gamma_\mu \gamma_5 d|0\rangle = i f_\pi  p_\mu,\\ \nn
&& \langle\eta(p)|\bar u \gamma_\mu \gamma_5 u|0\rangle = 
\langle\eta(p)|\bar d \gamma_\mu \gamma_5 d|0\rangle = 
- \frac{1}{2}\langle\eta(p)|\bar s \gamma_\mu \gamma_5 s|0\rangle = i \frac{f_\pi}{\sqrt{3}}  p_\mu,\\ \nn
&& \langle\rho^0(p)|\bar u \gamma_\mu u|0\rangle = 
- \langle\rho^0(p)|\bar d \gamma_\mu d|0\rangle = m_{\rho} f_{\rho}  \epsilon^*_\mu,\\ \nn
&& \langle\omega(p)|\bar u \gamma_\mu u|0\rangle = 
\langle\omega(p)|\bar d \gamma_\mu d|0\rangle = m_{\omega} f_{\omega}  \epsilon^*_\mu,
\ea
where $\epsilon_{\mu}$ is a vector meson polarization 4-vector; meson decay
constants and masses are $f_\pi=93$ MeV, $f_K=113$ MeV, $f_\rho=153$ MeV,
$f_{K^*}=224$ MeV, $f_{\omega} = 138$ MeV, 
$m_{\pi} = 139.6$ MeV, $m_K=494$ MeV, $m_{\eta} = 547$ MeV,  $m_{\rho} = 770$ MeV, 
$m_{\omega} = 782$ MeV, $m_{K^*} = 892$ MeV.

This channel-by-channel approach can not be extended beyond the threshold of $\eta'(985)$
meson production since its decay constant $f_{\eta'}$ is unknown. Above the $\eta'(985)$
threshold, in the domain (II),  
we approximate the decay rate of $\nu_h \rightarrow l(\nu) M$ by the decay rate of 
the inclusive process $\nu_h \rightarrow l(\nu) q_1 \bar q_2$ without 
perturbative and nonperturbative QCD corrections. This leading-order approximation should work 
for $m_{\nu h}>> \Lambda \approx 200$ MeV. At lower masses a more viable approach 
would be to relate the semileptonic $\nu_h$ decay rate by the dispersion relations 
to the imaginary parts of the W and Z self-energies $\Pi(s)$ in analogy to the approach
applied in the literature for the $\tau\rightarrow \nu + $ hadrons inclusive decay
\cite{SelfEnergy}.
However for our rough estimations we do not need this more sophisticated treatment and
will use the above mentioned leading-order approximation.  

The decay channels of the Dirac neutrino $\nu_h$ in the domain (I) are
\ba{channels}
&&(CC):\hspace*{1cm}\nu_h\longrightarrow \left\{
\begin{array}{l}
e^-e^+\nu_e,\
e^-\mu^+\nu_{\mu},\ \mu^-e^+\nu_e,\ \mu^-\mu^+\nu_{\mu},\\
e^-\pi^+,\ \ \mu^-\pi^+,\ \ e^-K^+,\ \ \mu^-K^+ , \\
e^-\rho^+,\ \ \mu^-\rho^{+},\ \ e^-K^{*+},\ \ \mu^-K^{*+}.
\end{array}\right.\\[3mm]
&&(NC):\hspace*{1cm}\nu_h\longrightarrow \left\{
\begin{array}{l}
\nu_i\ e^-e^+,\ \nu_i\ \mu^+\mu^-,\ \nu_i\ \pi^0,\ \\ 
\nu_i\ \eta,\ \nu_i\ \rho^0,\ \nu_i\ \omega,
\end{array}\right.
\ea
where $\nu_i= \nu_1, \nu_2, \nu_3$ are the three conventional light neutrino mass eigenstates. 
Latter on we neglect their masses since $m_{\nu i}<< m_h$.

Partial decay widths for these channels are readily calculated using the hadronic matrix 
elements in Eq. \rf{hadron}. The results can be summarized as
\ba{dec-width-4}
&&\Gamma(\nu_h\rightarrow l_1l_2\nu )= |U_{l_1 h}|^2
\frac{G_F^2}{192\pi^3} m_j^5 H(y_{l1},y_{l2}) \equiv |U_{l_1 h}|^2
\Gamma_3^{(l_1l_2)},\nn \\
&&\Gamma(\nu_h\rightarrow l P)= |U_{lh}|^2
\frac{G_F^2}{4\pi}f_{P}^2 m_j^3 F_P(y_l,y_{P})\equiv |U_{lj}|^2
\Gamma_2^{(lP)},\nn \\
&&\Gamma(\nu_h\rightarrow l V)= |U_{lh}|^2 \alpha_V
\frac{G_F^2}{8\pi}f_{V}^2 m_h^3 F_V(y_l,y_{V})\equiv |U_{lj}|^2
\Gamma_2^{(lV)},\\ \nn
&&\Gamma(\nu_h\rightarrow \nu_i l\bar l)= |\zeta_i|^2 \frac{m_h^5 G_F^2}{192 \pi^3}\ 
[a H_1^{(l)} + b H_2^{(l)}] \equiv |\zeta_i|^2 \Gamma^{(l)}_3,\\ \nn
&&\Gamma(\nu_h\rightarrow \nu_i P^0)= |\zeta_i|^2 \frac{m_h^3 G_F^2}{64 \pi}\ f_{\pi}^2 \ \alpha^{P^0}
(1 - y^2_{P^0})^2 \equiv |\zeta_i|^2 \Gamma^{(\nu P^0)}_2,\\ \nn
&&\Gamma(\nu_h\rightarrow \nu_i V^0)= |\zeta_i|^2 \frac{m_h^3 G_F^2}{16 \pi}\ f_{V^0}^2\  (1 - y^2_{V^0})^2
(1 + 2 y^2_{V^0}) \equiv |\zeta_i|^2 \Gamma^{(\nu V^0)}_2,
\ea
where 
\ba{def-12}
\zeta_i = \sum_{\alpha = e,\mu,\tau} U_{\alpha h} U_{\alpha i}^*
\ea
In Eq. \rf{dec-width-4} we denoted 
$P=\{\pi^+, K^+\}$, $V=\{\rho^+, K^{*+}\}$, $P^0 = \{\pi^0, \eta\}$, $V^0=\{\rho^0, \omega\}$, 
$y_i= m_{i}/m_h$, $\alpha^{\pi^0} = 1, \ \alpha^{\eta} = 1/3$, 
$a = 2 \sin^4\theta_W - \sin^2\theta_W +1/4,\ 
b= \sin^2\theta_W(\sin^2\theta_W -1/2)$. The kinematical functions are
\ba{kin-fun}
 &&H_P(x,y)= 12 \int\limits_{z_1}^{z_2} \frac{dz}{z}
(z-y^2)(1+x^2-z) \lambda^{1/2}(1,z,x^2)\lambda^{1/2}(0,y^2,z),\nn
\\ \nn
&& H_1^{(l)} = 12 \int\limits_w^1 \frac{d z}{\sqrt{z}} (z - 2 y_l^2)(1-z)^2 \sqrt{z - 4 y_l^2}, \\ \nn
&& H_2^{(l)} = 24 y_l^2 \int\limits_w^1 \frac{d z}{\sqrt{z}} (1-z)^2 \sqrt{z - 4 y_l^2},\\ \nn
&&F_P(x,y)= \lambda^{1/2}(1,x^2,y^2) [(1+x^2)(1+x^2-y^2) - 4 x^2],\nn \\ \nn
&&F_V(x,y)= \lambda^{1/2}(1,x^2,y^2) [(1-x^2)^2+(1+x^2)y^2) - 2 y^4],\\ \nn
\ea
where the integration limits are $z_1 = y^2,\ z_2=(1-x)^2$ and $ w = 4 y_l^2$.

In the mass domain (II) of Eq. \rf{parts} we approximate the CC and NC decay 
rates by the rates of the inclusive reactions 
\ba{NC-CC-incl}
(CC): \ \  \nu_h \longrightarrow l^- q_1 \bar q_2,\ \ \ \ \ \ \ \ \ 
(NC): \ \  \nu_h \longrightarrow \nu_i  q  \bar q, 
\ea
where $l = \{e,\mu\}$ and $q= \{u,d,s\}$. The corresponding leading order decay rate
formulas are
\ba{incl-formulas}
&&\Gamma(\nu_h\rightarrow l^- q_1 \bar{q_2})
= |U_{l h}|^2 \frac{G_F^2}{64\pi^3} m_h^5 ( |V_{ud}|^2+ |V_{us}|^2) \equiv 
|U_{l h}|^2 \Gamma^{(l X)},\\ \nn
&&\Gamma(\nu_h\rightarrow \nu_i\  q\  \bar{q}\ )
= |\eta_i|^2 \frac{G_F^2}{3072 \pi^3} m_h^5 [ 9 - 16 \sin^2\theta_W(1-\sin^2\theta_W)]
\equiv |\eta_i|^2 \Gamma^{(\nu X)}.
\ea
Here we neglected small $c\bar{d}, d\bar{c}$ contributions.

The following comment on the total NC decay rate $\Gamma_{\nu h}^{NC}$ is in order.
Summation over all the NC channels gives
\ba{NC-comment}
\Gamma_{\nu h}^{NC}&=& \Gamma_{\nu h}^{\nu} \sum_{i=1}^3 |\eta_i|^2 =
\Gamma_{\nu h}^{\nu} \sum_{\alpha,\beta = e,\mu,\tau} U_{\alpha h} U^*_{\beta h}
\sum_{i=1}^3 U^*_{\alpha i} U_{\beta i} =\\ \nn
&=&\Gamma_{\nu h}^{\nu} \sum_{\alpha,\beta = e,\mu,\tau} U_{\alpha h} U^*_{\beta h}
(\delta_{\alpha\beta} - U^*_{\alpha h} U_{\beta h}) \approx
\Gamma_{\nu h}^{\nu} \sum_{\alpha = e,\mu,\tau} |U_{\alpha h}|^2.
\ea
Here we assumed that there are only four neutrino mass eigenstates $\nu_{1,2,3}$, $\nu_h$
and used unitarity of the mixing matrix $U_{\alpha\beta}$. At the last step we 
neglected the subdominant term $\sim U^4_{\alpha h}$. Thus, 
both CC and NC decay rates are proportional to $\sim |U_{\alpha h}|^2$ 
and, as we mentioned at the beginning of this section, {\it a priori} 
are equally important. 

Taking into account this fact we collect the partial decay rates into the 
total $\nu_h$-decay width
\ba{total-4}
\Gamma_{\nu h} = 
|U_{e h}|^2\left(\Gamma_{\nu h}^{(e)} + \Gamma_{\nu h}^{(\nu)}\right) 
+ |U_{\mu h}|^2\left(\Gamma_{\nu h}^{(\mu)} + \Gamma_{\nu h}^{(\nu)}\right) +
|U_{\tau h}|^2 \Gamma_{\nu h}^{(\nu)}
\ea
with
\ba{NC-CC}
\Gamma_{\nu h}^{(l)} &=& (\Gamma_3^{( e \mu)}+ \Gamma_3^{(ll)})+
     \theta(m_{\eta'}-m_h)\sum_{M}
    \Gamma_2^{(l M)}+
  \theta(m_h-m_{\eta'})\Gamma^{(l X)},\\ 
\Gamma_{\nu h}^{(\nu)} &=& (\Gamma_3^{(e)}+ \Gamma_3^{(\mu)})+
     \theta(m_{\eta'}-m_h)\sum_{M^0}
    \Gamma_2^{(\nu M^0)}+
  \theta(m_h-m_{\eta'})\Gamma^{(\nu X)}  
\ea
where the summations run over $M=\pi^+,K^+,\rho^+,K^{*+}$ and $M^0 = \pi^0, \eta, \rho^0, \omega$.
  
If neutrinos $\nu_h$  are Majorana particles, i.e.  $\nu_h\equiv \nu_h^c$,
then both $\nu_h\rightarrow l^{-} X(\Delta L=0)$ and 
$\nu_h\rightarrow l^{+} X^c(\Delta L=2)$ decay channels are open. This results in 
multiplication of the right hand side of Eq. \rf{NC-CC} by factor 2.

\section{Bounds on heavy sterile neutrino masses and mixings}

Substituting the total $\nu_h$-decay width from Eq. \rf{total-4} into 
the resonant formula \rf{estim3}, we can derive, from the
experimental bounds  \rf{tdec-lim}, constraints on the $\nu_h$
neutrino mass $ m_{h}$ and the mixing matrix elements $U_{\alpha h}$.
In general these constraints represent a hardly readable 4-dimensional 
exclusion plot. However under certain simplifying assumptions one can 
infer more valuable information on the individual size of 
the mixing matrix elements.
In this paper we are interested  in the 
$U_{\tau h}$ matrix element which 
is not constrained in the literature in the $\nu_h$ mass range 
\rf{domain} (see \cite{PDG}). Recently the NOMAD collaboration\cite{NOMAD} 
obtained constraints for $m_h< 190$ MeV which overlap with a small part 
of this range.
Therefore, it would be interesting to infer individual constraints on this 
mixing matrix element, at least roughly.
A reasonable simplifying assumption would be to take $|U_{\tau h}|\sim |U_{\mu j}|\sim |U_{ej}|$.
Then from the experimental bounds \rf{tdec-lim} we derive a 2-dimensional 
$m_h-|U_{\tau h}|^2$ exclusion plot given in Fig.~2 for the case of Dirac $\nu_h$. 
Multiplying the Dirac case limiting curve in Fig.~2 by the factor 2 one obtains the 
exclusion plot for the case of Majorana $\nu_h$.
For comparison we also present in Fig.~2  the bounds from big-bang nucleosynthesis,
SN1987A \cite{Dolgov} and the NOMAD bounds \cite{NOMAD}. The cosmological bounds are 
shortly discussed in the next section. 
\begin{figure}[h!]
\hspace{-0.5 cm}
\mbox{\epsfxsize=14 cm\epsffile{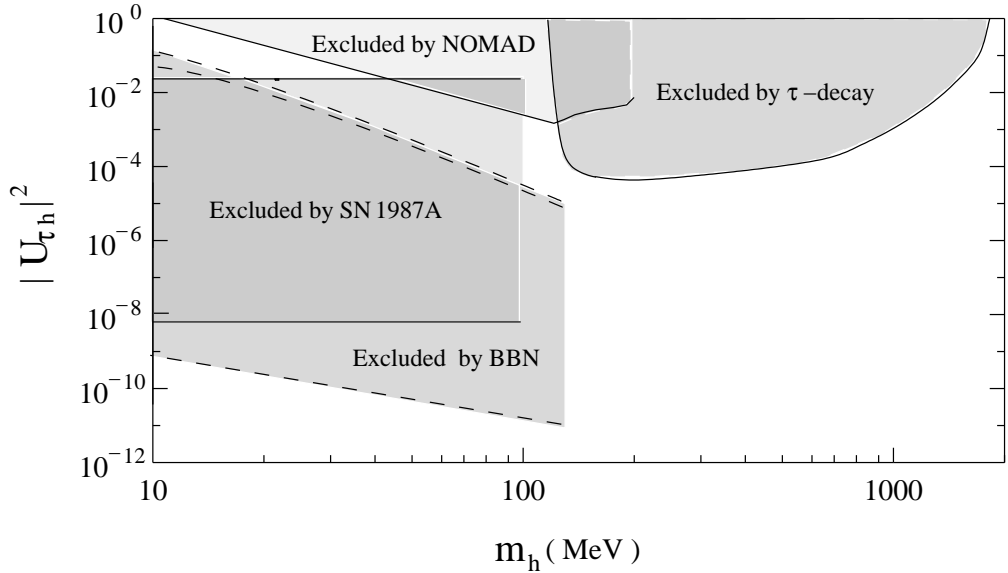}}
\caption{Exclusion plots in the plane $|U_{\tau h}|^2-m_h$.
Here $U_{\tau h}$ and $m_h$ are the heavy neutrino $\nu_h$
mixing matrix element to $\nu_{\tau}$ and its mass respectively.
The shaded regions are excluded by the big bang nucleosynthesis,
the duration of the SN 1987A neutrino burst \protect\cite{Dolgov},
by the NOMAD collaboration \protect\cite{NOMAD} and 
by $\tau \rightarrow e^{\pm}(\mu^{\pm})\pi^{\mp}\pi^-$ decay
[present result].
}
\vskip 1cm
\end{figure}

Note that at certain points of the mass range 
\rf{domain} the limits in Fig. 2 are incompatible with our 
assumption that $|U_{\tau h}|\sim|U_{\mu j}|\sim|U_{ej}|$ since 
other experiments already established limits like 
$|U_{e h}|^2, |U_{\mu h}|^2 < 10^{-8}$ at $m_h = 350, 400$ MeV.
Thus our limits around these points have to be treated  with caution. 
Let us note that all known constraints on $|U_{l h}|$ \cite{PDG} 
have been obtained under certain simplifying assumptions such as those that we used above. 

However, there is a special case when limits on $|U_{\tau h}|$ can be extracted without 
any simplifying assumptions. This case is realized when there are experimental upper 
bounds on the rates of three processes involving $\tau\tau$, $\tau\mu$ and 
$\tau e$ lepton pairs. These could be, for instance, decays
\ba{B-dec}
B^+_c \rightarrow \tau^+\tau^+ \pi^-,\ \tau^+\mu^+ \pi^-,\ \tau^+e^+ \pi^-.
\ea
Intersection of their resonant regions is $1917\ \mbox{MeV}\leq m_h \leq 4620$ MeV.
In this domain the decay rates can be written schematically as
\ba{Bc}\nn
&&\Gamma_{\tau\tau} = \frac{|U_{\tau h}|^4}{a |U_{e h}|^2 + b |U_{\mu h}|^2 + c |U_{\tau h}|^2},
\ \ \ 
\Gamma_{\tau\mu} = \frac{|U_{\tau h}|^2|U_{\mu h}|^2}{a |U_{e h}|^2 + b |U_{\mu h}|^2 + c |U_{\tau h}|^2},
\\ 
&&\hspace*{3.5cm} \Gamma_{\tau e} = \frac{|U_{\tau h}|^2|U_{e h}|^2}{a |U_{e h}|^2 + b |U_{\mu h}|^2 + c |U_{\tau h}|^2}.
\ea
From these Eqs. we find 
\ba{solution}
|U_{\tau h}|^2 = a \Gamma_{\tau e} + b \Gamma_{\tau\mu} +  c \Gamma_{\tau\tau}.
\ea
Notice that all the decay rates appear with positive degree. 
Only in such a case we can set upper bound on $|U_{\tau h}|^2$ 
having upper experimental bounds $\Gamma_{\tau l}\leq Exp_l$.
This would not be the case, for instance, in the system 
$\Gamma_{\tau\mu}, \ \Gamma_{\tau e}, \ \Gamma_{e e}$.

At present the described approach can not be realized on practice due to 
absence of necessary experimental data. However the required 
upper limits on the semileptonic $B_c$-decays can be, probably, obtained 
in future experiments at the B-factories. 
Then, derivation of limits on $|U_{\tau h}|$ free of simplifying {\it ad hoc} 
assumptions will become possible.

\section{Heavy sterile neutrinos in astrophysics and cosmology}

It is well known that massive neutrinos may have
important cosmological and astrophysical implications.
They are expected to contribute to the mass density of
the universe, participate in cosmic structure formation,
big-bang nucleosynthesis, supernova explosions, imprint themselves
in the cosmic microwave background etc.\ (for a review see, for instance,
Ref.  \cite{Raffelt}). This implies certain constrains on the neutrino masses
and mixings.
Currently, for massive neutrinos in the mass region
~\rf{domain}, the only available cosmological
constraints arise from the mass density of the
universe and cosmic structure formation.

The contribution of stable massive neutrinos to the mass density
of the universe is described by the ``Lee-Weinberg" $\Omega_{\nu}
h^2-m_{\nu}$ curve. From the requirement that the universe is not
``overclosed" this leads to the two well know solutions $m_{\nu}\
\leq 40$eV and $m_{\nu} \ \gsim 10$GeV which seem to exclude the
domain Eq. \rf{domain}. However for unstable neutrinos the
situation is different.  They may decay early to light particles
and, therefore, their total energy can be significantly
``redshifted" down to the ``overclosing" limit. Constraints on the
neutrino life times $\tau_{\nu_j}$ and masses $m_j$ in this
scenario are found in Ref. \cite{DM}. In the mass region
\rf{domain} we have an order of magnitude estimate
\ba{DM}
\tau_{\nu_j}< (\sim 10^{14}) \mbox{sec} \ \ \ \ \ \mbox{Mass Density limit}
\ea

Decaying massive neutrinos may also have specific impact on
the cosmic structure formation introducing new stages
in the evolution of the universe. After they decay into light relativistic
particles the universe returns for a while from the matter to the radiation
domination phase. This may change the resulting density fluctuation spectrum since
the primordial fluctuations grow due to gravitation instability during the matter
dominated stages.
Comparison with observations leads to an upper bound on
the neutrino life time \cite{StructForm}. In the mass region \rf{domain}
one finds roughly
\ba{SF}
 \tau_{\nu_j}< (\sim 10^{7})\mbox{sec} \ \ \  \mbox{Structure Formation limit}
\ea
On the other hand, from our formula \rf{total-4} we estimate 
\ba{theor}
\mbox{few sec} < \tau_{\nu_j}, \ \ \ \ \mbox{Theoretical limit.}
\ea
Thus massive neutrinos with masses in the interval \rf{domain} are not yet
excluded by the known cosmological constraints \rf{DM}, \rf{SF}.

Big-bang nucleosynthesis(BBN) and the SN 1987A neutrino signal may presumably
lead to much more restrictive constraints \cite{Dolgov}.
The impact of heavy sterile neutrinos $\nu_h$ on the BBN is twofold: via their contribution to 
the cosmic energy density and via the secondary light neutrinos $\nu_e, \nu_{\mu}, \nu_{\tau}$ 
produced in $\nu_h$-decays. The first issue results in faster expansion which enhances
the neutron-to-proton ration $n/p$. The secondary light neutrinos can produce 
also an opposite effect. The increased number of $\nu_e$ maintains 
neutron-proton thermal equilibrium longer 
diminishing the $n/p$-ratio at the freeze-out point. There are also effects related
to the $\nu_e$ energy spectrum distortion and energy injection to the electromagnetic radiation 
from the $e^+e^-$ pairs produced in $\nu_h$-decays \cite{Dolgov}.
The observed duration of the neutrino signal from the SN 1987A stringently constraints 
the energy emitted from the core in the form of penetrating particles like neutrinos. 
The sterile neutrinos could affect this energy-loss limits and, therefore, their 
mixings with the active neutrinos have to obey certain constraints. 

Analysis of the BBN and SN 1987A constraints on $\nu_h$ parameters has been made 
in Ref. \cite{Dolgov}. We present the corresponding exclusion plots
in Fig. 2 for comparison with our limits. 
Unfortunately, the analysis of Ref. \cite{Dolgov} is restricted to the mass 
range $m_h < 200$ MeV for the BBN and $m_h < 100$ MeV for the SN 1987A case.
From the general appearance of the excluded bends one may expect that    
the extension of the SN 1987A bend can overlap with the region excluded by the $\tau$-decay,
producing very stringent combined constraint around $|U_{\tau h}|^2<10^{-8}$. Even stronger
constraint would appear in the domain where the SN 1987A bend simultaneously overlaps 
with the BBN excluded bend. In this case a combined constraint would be, probably, more 
stringent than $|U_{\tau h}|^2<10^{-12}$. 

\section{Conclusion}

We discussed a generic model with sterile neutrinos which can contain heavy $\nu_i$
mass eigenstates. If among them there are $\nu_h$ states with masses
$ 140.5 \mbox{MeV} \leq m_{h} \leq 1637 \mbox{MeV}$ they can resonantly contribute 
to $\tau^-\rightarrow e^{\pm}(\mu^{\pm})\pi^{\mp}\pi^-$ decays. 

We derived decay rates for these processes valid both outside and inside of the resonant 
$m_h$ region \rf{domain}. In the latter case knowledge of the total $\nu_h$-decay width
$\Gamma_{\nu h}$ is required. We calculated the $\Gamma_{\nu h}$ in the whole resonant $m_h$ 
region \rf{domain} taking into account both charged and neutral current decay channels. 

The effect of resonant enhancement of the heavy sterile neutrino $\nu_h$ contribution 
allowed us to extract from the experimental upper bounds on the $\tau$-decay rates  
rather stringent constraints on $\nu_{\tau}-\nu_h$ mixing matrix
element $|U_{\tau h}|$ in the whole resonant region \rf{domain}. 
The corresponding exclusion plot is presented in Fig. 2 together with 
the recent constraints from NOMAD collaboration \cite{NOMAD} and cosmological constraints
\cite{Dolgov}. 
The $\tau$-decay constraints were derived under certain simplifying assumptions 
about $\nu_h$-mixing with the other $\nu_{\alpha}$ flavors. These or similar 
assumptions are always required when $\nu_h$-mixing with one selected $\nu_{\alpha}$ flavor 
is extracted from those processes which have been studied in the literature. We propose  
the set of processes 
$B_c\rightarrow \tau\tau\pi, \tau\mu\pi, \tau e\pi, \mu\mu\pi, ee\pi$ which would 
allow one to avoid such {\it ad hoc}  assumptions and extract individual constraints 
on $|U_{\tau h}|, |U_{\mu h}|, |U_{e h}|$. 
Finally, we discussed cosmological implications of the heavy sterile neutrinos $\nu_h$ and 
pointed out that extension of the known big bang nucleosynthesis and the SN 1987A 
constraints \cite{Dolgov} to the resonant region \rf{domain} could  
significantly strengthen our constraints on $|U_{\tau h}|$ 
by combining them with these cosmological constraints.

\vskip10mm
\centerline{\bf Acknowledgments}
We thank Ya. Burdanov, G. Cveti\v c, C. Dib and O. Espinosa for 
fruitful discussions. We are grateful to S. Gninenko for helpful comments 
on the recent NOMAD results. This work was supported in part by Fondecyt (Chile) under 
grant 8000017, by a C\'atedra Presidencial (Chile) and by RFBR (Russia) under 
grant 00-02-17587. 
\bigskip
\end{document}